\documentclass{emulateapj}
\usepackage[dvips]{color}

\shorttitle{Burst oscillations in HETE J1900.1-2455}
\shortauthors{Watts et al.}

\begin{document}

\title{Discovery of burst oscillations in the intermittent accretion-powered millisecond pulsar HETE J1900.1-2455}

\author{Anna L. Watts\altaffilmark{1}, Diego Altamirano, Manuel Linares,
  Alessandro Patruno, Piergiorgio
  Casella, Yuri Cavecchi, Nathalie Degenaar, Nanda Rea, Paolo Soleri, Michiel van der Klis \&
Rudy Wijnands} 
\affil{Astronomical Institute ``Anton Pannekoek'', University of
  Amsterdam, Kruislaan 403, 1098 SJ Amsterdam, the Netherlands}
\altaffiltext{1}{A.L.Watts@uva.nl}

\begin{abstract}
We report the discovery of burst oscillations from the intermittent
accretion-powered millisecond pulsar (AMP) HETE J1900.1-2455, with a
frequency $\sim 1$ Hz below the known spin
frequency.  The burst oscillation properties
are far more similar to those of the non-AMPs and Aql X-1 (an 
intermittent AMP with a far lower duty cycle), than those of the AMPs SAX J1808.4-3658
and XTE J1814-338.  We discuss the implications for models of the
burst oscillation and intermittency mechanisms.    
\end{abstract}


\section{Introduction}

There are two ways in which we can measure the spin of rapidly
rotating accreting neutron stars: via accretion-powered pulsations (due to magnetic channeling of infalling material), or burst
oscillations (coherent pulsations seen during the thermonuclear
explosions of accreted fuel that give rise to Type I X-ray bursts).
The mechanism that generates the brightness asymmetries responsible for burst
oscillations is not yet known.  Once understood,
however, it should shed light on the thermonuclear burning process and the composition of the outer
layers of the neutron star \citep{str06, wat08b}. 

Burst oscillations often drift upwards in frequency by a few Hz during
a burst.  However the stability of the asymptotic
frequency for a particular source \citep{mun02}, and the detection of the same frequencies
in superbursts as well as Type I bursts \citep{str02}, suggested even
at an early stage that burst oscillation frequency might be (to within a few Hz) the
spin frequency.  This was
confirmed in 2003 with the discovery of burst oscillations at or
within a few Hz of the spin frequency in the accretion-powered
millisecond pulsars (AMPs)
SAX J1808.4-3658 and XTE J1814-338 \citep{cha03, str03}.  
The near identity of burst oscillation frequency and spin frequency is
now the major constraint on models for the burst oscillation
mechanism.  

The AMPs are particularly useful when investigating the burst oscillation
mechanism, since they are the only sources in which we can gauge the
effects of the magnetic field and asymmetric fuel
deposition.  They are also the only stars where we can measure the
size and sign of the small offset
between the spin and burst oscillation frequency. However both SAX J1808.4-3658 and XTE J1814-338 have
burst 
oscillations with quite atypical properties compared to the other
sources (Section \ref{disc}).  It is not yet clear whether we are
seeing a 
continuum of behavior (which could be explained by one burst
oscillation mechanism) or completely separate classes.  

The picture has evolved in the last three years with the
discovery of three intermittent AMPs:  HETE J1900.1-2455, SAX J1748.9-2021 and Aql X-1
\citep{kaa06, gal07, gav07, alt08, cas08}. Unlike the other AMPs
these sources show accretion-powered pulsations only
sporadically.  What cause the intermittency is also not known, with
models including sporadic
obscuration of the magnetic poles (\citealt{gog07} and references therein),
wandering of the accretion funnel and hence the hot spot around the magnetic 
pole \citep{rom03, rom04, lam08}, or magnetic field burial
\citep{cum01, cum08}.  

If the magnetic field is what makes the burst oscillations of the AMPs
so different from those of the other stars, then an intermittency
mechanism that affects the magnetic field may also influence burst
oscillation properties.  Burst oscillations from the intermittent AMPs
(all of which burst) are therefore an important piece of the puzzle.   Until now, however,
only Aql X-1 has shown burst oscillations, at $\approx 0.5$ Hz below
the spin frequency \citep{zha98, cas08}.  The properties of the Aql X-1 burst oscillations are very similar to
those of the non-AMPs:  however this source has 
the lowest pulsation duty cycle of any of the intermittent AMPs.  In this
Letter we report the discovery of burst oscillations from the
intermittent AMP HETE J1900.1-2455 \citep{wat09}, and consider the
implications for the burst oscillation and intermittency mechanisms.

\section{Burst analysis}
\label{data}

HETE J1900.1-2455 was first detected in June 2005, and has remained in
outburst ever since apart from a 3 week period of quiescence in 2007
\citep{deg07}.  It was quickly identified as an 
AMP with a spin frequency of 377.3 Hz, an orbital period of 83.3
minutes, and a low  ($< 0.1 M_\odot$) mass companion  \citep{kaa06}.
It is at a distance of $\approx 5$ kpc, determined from Eddington limited X-ray bursts
\citep{gal08a}.  Accretion-powered pulsations of low
fractional amplitude ($\lesssim 3$ \%) were detected sporadically during the first two months
of outburst, but have not been seen since \citep{gal07, gal08b}. There
is some evidence that the
intermittent appearance of the accretion-powered pulsations may be related to the occurrence of Type I X-ray bursts, but
whether the relationship is causal is not clear \citep{gal07}. From
2005-2008 several X-ray bursts from this source were detected by {\it
  HETE-II}, the {\it Rossi X-ray Timing Explorer} (RXTE)
and SWIFT\footnote{\citet{gal08b} list most of
  the bursts detected during 2005-2008.  One
  additional burst during this period was recorded by RXTE on December 5 2007 (05:57
  UTC), and there is a candidate burst in slew data on
  November 27 2007.}.  We searched the bursts detected by both the RXTE Proportional Counter
Array (PCA) and the SWIFT Burst Alert Telescope (BAT)
for oscillations, but found no significant signal.   

\subsection{Burst oscillations}
\label{data1}

During routine monitoring on April 2 2009 (08:57 UTC),
RXTE detected another X-ray burst with both the PCA
and the {\it High Energy X-ray Timing Experiment} (HEXTE).  Timing
analysis was conducted using 125 $\mu$s time resolution PCA event mode
data from the two active PCUs (PCUs 0 and 2).  We used all photons in the 2-30 keV range, the
band where burst emission exceeds the persistent level.  The data were
barycentered using the JPL DE405 and spacecraft ephemerides, with
the source position of \citet{fox05}.  Some event mode data
overruns occurred in the burst peak, leading to short data gaps.

\begin{figure}
\centering
\includegraphics[width=8.5cm, clip]{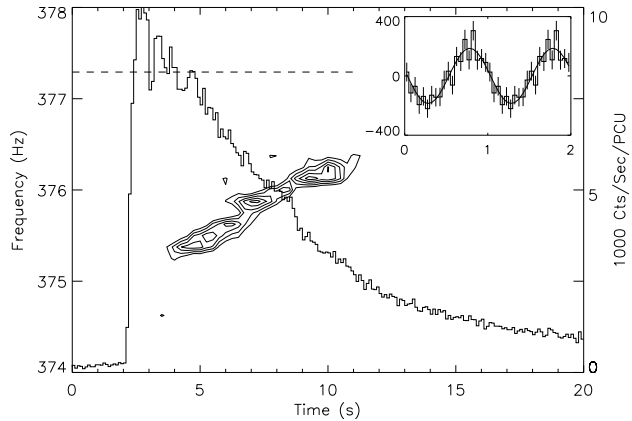}
\caption{Lightcurve and dynamical power spectrum (2-30 keV) for the April 2 2009
  burst. The dynamical power spectrum uses overlapping 4 s bins, with new bins starting at 0.25 s 
intervals.  We use a Nyquist frequency of 2048 Hz and an interbin
response function to reduce artificial drops in amplitude as the
frequency drifts between
Fourier bins \citep{vand89}. The contours show
Leahy normalized powers of 20-40, increasing in steps of 5. The dashed
line indicates the 377.3 Hz spin frequency. The inset shows the
pulse profile of the burst oscillations for the 8s interval where they
are detected, folded with the best fit polynomial frequency model
(counts per 0.05 cycle phase bin against phase in cycles), with the mean subtracted. The resulting profile is well
fit by a simple sinusoid (overplotted) with no requirement for any
harmonic content.  Two cycles are shown for clarity.   }
\label{dps}
\end{figure}

A dynamical power spectrum (Figure \ref{dps}) reveals strong burst oscillations during
the initial decay of the burst, drifting upwards by about 1 Hz.  Selecting only
data after t=4 s in Figure \ref{dps} (the time of the final data gap),
we find maximum Leahy
power in the range 17-30 in 4 independent consecutive 2 s bins, an
extremely robust detection.  The oscillations fall below the detectability threshold when the
frequency is 376.3 Hz, 1 Hz below the spin frequency.  Although the
binary ephemeris cannot be extended to 2009 (the errors are too large), orbital
Doppler effects will shift the spin frequency by 0.009 Hz at
most \citep{kaa06}, so the offset in the frequencies is secure. A search for
accretion-powered pulsations during the observation revealed no
significant signals.

To study the pulse profile, we extracted data for the 8s
period (t = 4-12s in Figure \ref{dps}) during
which the oscillations are detected.  We fitted the frequency drift
using a polynomial model to maximise the power.  Power was maximised
with a quadratic drift model, $\nu = \nu_0 +
\dot{\nu}_0 (t-t_0) + \ddot{\nu}_0(t-t_0)^2$, which yielded a
Leahy normalised power of 88 (the best linear model gave a maximum
power of 78, and higher order polynomial terms yielded no further improvement).  Using the best fit
quadratic frequency model we generated a folded profile.  The persistent
(non-burst) emission level was estimated from the
$\sim 1000$s prior to the burst.  The resulting pulse profile is shown
inset in Figure \ref{dps}. The pulse is well fit with a simple sinusoid
($\chi^2$/{\it dof} = 31/37), with no need for
harmonic content.  The fractional amplitude was 
($3.5 \pm 0.3$)\% Root Mean Square (RMS) (2-30 keV), comparable to the fractional
amplitude of the accretion-powered pulsations of this source.  The
amplitude of the pulsations rises with energy, from ($3.1 \pm 0.5$) \% RMS
in the 2-5 keV band to ($5.5\pm 0.6$)\% RMS in the 10-20 keV band.
The burst oscillations also show hard lags, with the 10-20 keV pulse
lagging the 2-5 keV pulse by (0.09 $\pm$ 0.03) cycles.   
\newline

\subsection{Burst properties}
\label{data2} 

To study the characteristics of the burst containing the oscillations, we extracted spectra every
0.25s from the PCA Event data (E\_125us\_64M\_0\_1s), which has 64
energy channels between 2 and 60 keV. We used data from PCUs 0
and 2, which were operating at the time of the X-ray burst.  We
generated an instrument response matrix for each spectrum, and 
fitted it using XSPEC version 11.3.2. We added a 1\% systematic error to
the spectra and restricted the spectral fits to the energy range
2-25 keV. We extracted spectra using 100s segments before and
after the burst, and used them as 
background in our fits. We fitted each of the 0.25s spectra with
an absorbed ({\tt wabs}, \citealt{mor83}) blackbody model
({\tt bbodyrad} model in XSPEC). This approach, which takes
account of the contribution of the persistent and background
emission, is standard procedure in X-ray burst analysis
(see for example \citealt{kuu03}).

The simple blackbody fit described above yields a peak luminosity of $3.6 \times 10^{38}$ ergs
s$^{-1}$, and a total energy of $2.7 \times 10^{39}$ ergs.  The
evolution of the temperature $T_{bb}$ and the radius $R_{bb}$ 
given by the spectral fits resembles those seen in
photospheric radius expansion (PRE) bursts \citep{gal08a},
i.e. (i) $R_{bb}$ reaches a local maximum close to the time of the
peak flux, (ii) lower values of $R_{bb}$ were measured following the
flux maximum and (iii) there is a local minimum in
$T_{bb}$ at the same time as the maximum in $R_{bb}$.  This conclusion
(and the derived peak flux and fluence)
should be treated with care, however, since the $\chi^2/dof$
of the spectral fits can reach values as high as 4 or
5. The problem stems from an excess at high energies.  Similar issues
were reported by \citet{gal08b} for other 
bursts from this source, and the physical cause is not yet understood.  Adding a
power law to our model, for example,
reduces $\chi^2/dof$ to below 2: the evolution of $T_{bb}$ and $R_{bb}$
in this case still suggests a PRE burst.  Photospheric touchdown occurs $3.5-5$s after the
start of the burst, consistent with the first detection of the
burst oscillations. 

\begin{figure}
\centering
\includegraphics[height=8.5cm, angle=270, clip]{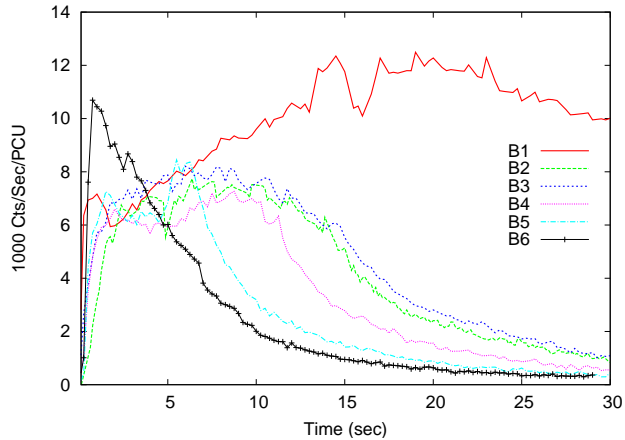}
\caption{2-25 keV lightcurves for the bursts recorded by the RXTE PCA from
  2005-2009. Background and
  persistent emission have been subtracted and deadtime corrections
  applied (this was not done in Figure \ref{dps}).  The bursts are numbered according to their position in
  the color-color diagram (Figure \ref{ccd}).  B6, the burst with
  the oscillations, has quite a different lightcurve.  }
\label{blc}
\end{figure}

The parameters of the April 2 burst do not differ dramatically from
those of other bursts from HETE J1900, which also show evidence for
PRE subject to uncertainties about the spectral fits
\citep{gal08a,gal08b}. The burst with oscillations does not however
have the 
extended or double peak structure exhibited by most of the 
other bursts (Figure \ref{blc}). The burst lightcurve shows a fast rise, $\approx 0.4$
s defined as in \citet{gal08a}.  The decay
can be modelled with a double exponential with decay timescales as
of 7.3 s and 8.4 s respectively.  Total burst duration is $\approx 60$ s (burst end
is defined as the time when flux falls to 1\% of the peak flux,
corrected for persistent emission).

\subsection{Persistent emission}
\label{data3}

\begin{figure}
\centering
\includegraphics[height=8.5cm,angle=270]{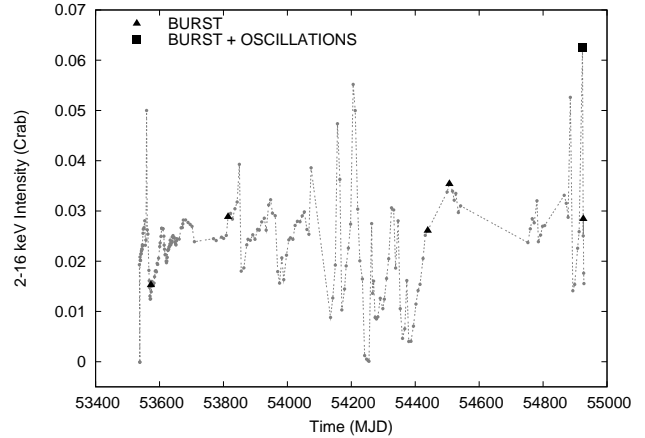}
\caption{Average 2-16 keV intensity per RXTE observation for HETE
  J1900.1-2455 from 2005 to April 4 2009 (data from
March-October 2008 are not yet public).  Observations containing bursts are marked in black:
triangles (no oscillations), square (oscillations). We include the
four confirmed bursts detected by the PCA from 2005-2008 (the
candidate slew burst also occurs in the hard state, but is not shown), the burst
with oscillations from April 2 2009 and an additional burst detected
by the PCA on April 4 2009.}
\label{lc}
\end{figure}

\begin{figure}
\centering
\includegraphics[height=8.5cm,angle=270, clip]{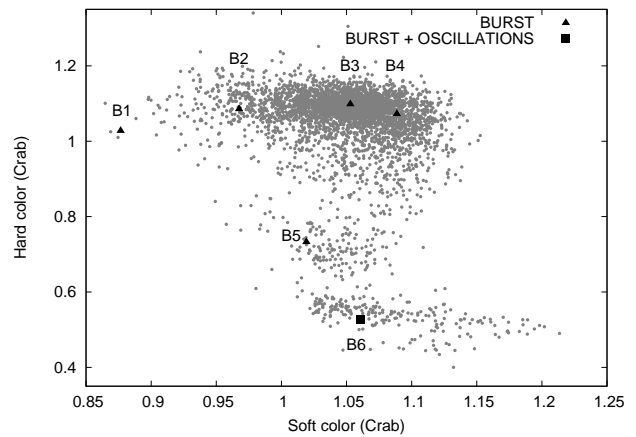}
\caption{Color-color diagram for HETE J1900.1-2455 using all publicly available
  RXTE data from 2005 to April 4 2009 (excluding observations with intensities
  below 2 mCrab, see Figure \ref{lc}). The
  source behaves like a typical atoll source \citep{has89}.  The grey
  dots show the colors computed from 160s segments of data. Observations containing bursts are marked in black: triangles (no oscillations), square
  (oscillations).  Lightcurves for the bursts are shown in Figure \ref{blc}.} 
\label{ccd}
\end{figure}

Figure \ref{lc} shows the intensity, defined
as the countrate in the 2.0--16.0 keV band, for all publicly available
observations from 2005-2009. Figure \ref{ccd} shows the corresponding
color-color diagram.  Data were deadtime corrected, background
subtracted, and X-ray bursts removed.  For each observation  we calculated X-ray colors from the Standard2
data.  We defined soft color as the ratio between count rates in the
3.5--6.0 and 2.0--3.5 keV bands and hard color as the ratio between
count rates in the 9.7--16.0 and 6.0--9.7 keV bands. We
normalized colors and intensity to the Crab values nearest 
in time \citep{kuu94} and in the same PCA gain epoch (see for example
\citealt{vans03}).  

It is clear from Figures \ref{lc}-\ref{ccd} that the observation
where the burst oscillations were detected was rather unusual.  The
burst happened when source intensity was at its highest recorded level ($\approx 63$
mCrab) and when it was in the soft (banana) state.  All previous bursts
have been detected in harder states.  Fitting the PCA spectrum with an absorbed disk-blackbody plus
power-law model, and assuming a standard bolometric correction factor
of 2 \citep{int07} we find an unabsorbed bolometric flux of $3.5 \times 10^{-9}$ ergs s$^{-1}$
cm$^{-2}$ (interstellar absorption was fixed to $1.6 \times 10^{21}$ cm $^{-2}$). At
  a distance of 5 kpc, this corresponds to 4\% of the Eddington
  luminosity if we assume $L_\mathrm{Edd}=2.5 \times 10^{38}$ erg
  s$^{-1}$.

\section{Discussion}
\label{disc}

What causes burst oscillations is still not understood. Flame spread
from a point should lead to asymmetries in the early phase of the
burst.  However, while this may explain the presence of burst
oscillations during 
the rise, it is not sufficient to explain the continuing
presence of large-scale asymmetries in the burst tail once the entire
stellar surface has ignited \citep{str06}.  This has led to the
consideration of alternative models.  Non-radial global
oscillations in the surface layers of the star, excited by the flame
spread, remain a promising possibility \citep{hey04}.
In this case the brightness asymmetry would be caused by variations in
ocean height associated with the mode. Inertial frame pattern speed can
be very close to the spin rate, and cooling of the layers in the
aftermath of the burst would naturally lead to frequency drift.   The
major problem of this model is that it overpredicts the size of the
frequency drift compared to observations \citep{pir05, ber08}.  Alternative possibilities
like photospheric modes \citep{hey04} or shear oscillations \citep{cum05}
also have shortcomings in their present form \citep{ber08}.  Current
efforts to resolve these problems are focusing on the role of the
magnetic field, which can dominate the dynamics in the surface layers
and is expected to have a large effect on surface modes.  

If the magnetic field is the most important factor determining the properties of
burst oscillations then this might lead to a natural explanation for the
differences in the properties of the burst oscillations of 
SAX J1808.4-3658 and XTE J1814-338 (hereafter collectively referred to
as SX: \citealt{cha03, str03, wat05, wat06}) compared to those
of the non-AMPs \citep{mun02, mun02b, mun03, mun04, gal08a}.
These differences can be summarized as follows.  (1) SX show
oscillations in all of their bursts, even in the hard state; the non-AMPs
do not (oscillations are detected primarily although not exclusively in the
soft state). (2) SX show oscillations throughout their bursts (except
during PRE); the non-AMPs, in most cases, do not. (3) SX burst oscillations have
amplitudes that fall with energy; the non-AMP oscillations have
amplitudes that rise. (4) SX burst oscillations have detectable
harmonic content; the non-AMPs have none. (5) Frequency drift in SX is
either very fast or non-existent; in the non-AMPs it is much slower.

The magnetic field also plays an important role in models for
intermittency of accretion-powered pulsations.  In the obscuration
and accretion stream wander models, the field is always present at the
level necessary to channel the flow:  but the accretion hot spot either wanders
out of the line of sight or is obscured by
magnetospheric material.   In the magnetic burial model the field
strength and geometry change as the field is suppressed by accretion.
If magnetic field affects both intermittency and burst oscillations,
then studying the latter may enable us to pinpoint the cause of the
former.  

In Aql X-1 the burst oscillations are similar in properties to those
of the non-AMPs:  however this source has an exceptionally low duty
cycle so its intermittent pulsation episode may have been triggered by
an extremely rare event.  In HETE J1900.1-2455 the accretion-powered
pulsations lasted for much longer.   Its burst oscillations, however,
also behave like those of the non-AMPs.  This has important implications.
If hot spot wander or obscuration models for intermittency are correct,
then a field strong enough to channel infalling material cannot be the
only factor causing the atypical burst oscillations of SAX J1808.4-3658
and XTE J1814-338.  Some other factor, such as the presence of a
strong temperature gradient around the magnetic pole,
must also play a role in the burst oscillation mechanism
\citep{wat08c}.  If on the other hand the field has been buried in
HETE J1900.1-2455, then it must be screened to a depth where it is unable
to affect the burst oscillation mechanism on the timescale of the burst.   Detailed calculations
will be required to resolve this issue, but if screening at the burning depth is not
possible, then this could point to photospheric rather than
oceanic modes as a cause of burst oscillations.

One other point of note is that in both Aql X-1 and HETE J1900.1-2455,
burst oscillation frequency remains below spin frequency.  Given the
similarities in properties it seems probable that this is the case for
non-AMPs as well.  However it is not
the case for SAX J1808.4-3658 and XTE J1814-338.  In SAX J1808.4-3658 the burst oscillations
are first detected below the spin frequency but rapidly overshoot it,
settling $\sim$ 0.1 Hz above the spin frequency in the burst tail
\citep{cha03}.  In XTE J1814-338 the two frequencies are identical,
and in fact burst oscillations and accretion-powered pulsations are 
coherent and phase-locked \citep{str03, wat05, wat08c}.  Any unified
model of burst oscillations must be able to explain this diversity in the
relationship between burst oscillation frequency and spin. 


\end{document}